\documentclass[twocolumn, twocolappendix]{aastex631}
\usepackage{amsmath}
\usepackage{wrapfig}
\usepackage{svg}

\usepackage{amsmath}
\usepackage{amssymb}
\usepackage{epsfig}
\usepackage{graphicx}
\usepackage{subfigure}
\usepackage{xcolor}
\usepackage{float}
\usepackage{epstopdf}
\usepackage{natbib}
\usepackage{txfonts}
\usepackage{mathrsfs}
\usepackage{graphics}
\usepackage{keyval}
\usepackage{url}
\usepackage{multirow}
\usepackage{ulem}
\usepackage{bm}
\usepackage{dutchcal}
\usepackage{lipsum}
\usepackage{orcidlink}

\newcommand{\Rmnum}[1]{\expandafter\@slowromancap\romannumeral #1@}
\allowdisplaybreaks[3]

\newcommand \beq{\begin{equation}}
\newcommand \eeq{\end{equation}}
\newcommand \bey{\begin{eqnarray}}
\newcommand \eey{\end{eqnarray}}
\newcommand \Myr{\, {\rm Myr} }
\newcommand \Gyr{\, {\rm Gyr} }

\newcommand \pc{\, {\rm pc} }

\newcommand \kpc{\, {\rm kpc} }

\newcommand \msun{{\rm M}_\odot}
\newcommand \Msun{{\rm M}_\odot}

\newcommand \kms{\, {\rm km \, s}^{-1} }
\newcommand \kmssq{\, {\rm km}^{2} \, {\rm s}^{-2} }

\newcommand \vc{\bm{v}_{\rm c}}
\newcommand \rc{\bm{r}_{\rm c}}
\newcommand \ri{\bm{r}_{i}}
\newcommand \ristroke{\bm{r}_{i}'}

\begin{document}

\title[Energy evolution in shell-galaxy-progenitor]{Energy evolution in the progenitor of galaxy shells: a semi-analytical model}

\author{Beibei Guo ~\orcidlink{0000-0003-3390-0438}}
\affiliation{CAS Key Laboratory for Research in Galaxies and Cosmology, Department of Astronomy, University of Science and Technology of China, Hefei 230026, China}
\affiliation{School of Astronomy and Space Science, University of Science and Technology of China, Hefei 230026, China}
\author{Xufen Wu~\orcidlink{0000-0002-1378-8082}}\altaffiliation{Corresponding author: xufenwu@ustc.edu.cn}
\affiliation{CAS Key Laboratory for Research in Galaxies and Cosmology, Department of Astronomy, University of Science and Technology of China, Hefei 230026, China}
\affiliation{School of Astronomy and Space Science, University of Science and Technology of China, Hefei 230026, China}
\author{HongSheng Zhao~\orcidlink{0000-0001-9715-8688}}
\affiliation{CAS Key Laboratory for Research in Galaxies and Cosmology, Department of Astronomy, University of Science and Technology of China, Hefei 230026, China}
\affiliation{School of Physics and Astronomy, University of St Andrews, North Haugh, Fife, KY16 9SS, UK}
\author{Lulu Fan~\orcidlink{0000-0003-4200-4432}}\altaffiliation{Corresponding author: llfan@ustc.edu.cn}
\affiliation{CAS Key Laboratory for Research in Galaxies and Cosmology, Department of Astronomy, University of Science and Technology of China, Hefei 230026, China}
\affiliation{School of Astronomy and Space Science, University of Science and Technology of China, Hefei 230026, China}
\affiliation{Deep Space Exploration Laboratory, Hefei 230088, China}

\begin{abstract}
The stellar shells surrounding an elliptical galaxy, as remnants of a dwarf galaxy disrupted during merging, reveal the distribution of energy and angular momentum of the progenitor dwarf galaxy. We develop a semi-analytical model to describe the changes of energy $\Delta E_i$ and angular momentum $\Delta Lz_i$ for particles during the first infall. We show that these changes, induced by the self-gravity of the progenitor, are important in broadening the initial energy distribution of the Plummer or Hernquist progenitor model. Consequently, these changes are crucial in shaping the shells. In the free fall stage following the disintegration of the progenitor potential, particles are no longer bound by self-gravity but move within the gravitational potential of the target galaxy. We investigate the relationship between the radial period and the energy of particles undergoing radial motion. We show that an accurate model of the energy range of the dwarf galaxy at disruption is essential to predict the number of observable shells. 
\end{abstract}


\section{Introduction}
Deep imaging and the high-resolution cosmological simulations reveal the details of faint structures in shell galaxies \citep{Pop2018,Donlon2020,Alabi2020,Bilek2020}. The shell galaxies, first discovered by \citet{Arp1966}, are characterized by delicate, open concentric non-crossing arc structures which composed of stars. In the local Universe, approximately $10\%$ of elliptical or lenticular galaxies exhibit luminous sharp-edged stellar shells \citep{Malin1983,Schweizer1988,Colbert2001,Atkinson2013,bilek2016}. Stellar shells are found at varying distances from the centers of their host galaxies, ranging from several $\kpc$ to about $100~\kpc$. Shell galaxies are categorized by their morphologies into three types \citep{Prieur1990}. A Type I shell galaxy consists axially symmetric shells staggered around the galactic centers on both sides. A Type II shell galaxy, on the other hand, is a nearly circular galaxy in shape, with randomly distributed arcs around it. Lastly, a Type III shell galaxy contains irregular and complex arc structures. 

There has been extensive research on the formation mechanisms of shell structures since the 1980s, such as gas dynamical processes\citep{Fabian1980,Williams1985}, weak gravitational interaction models\citep{Thomson1990}, and the widely accepted merger models\citep{Schweizer1980,Quinn1984}. The idea that the formation of shells results from galaxy mergers originated from \citet{Schweizer1980}. Later, the merger model was developed by \citet{Quinn1984}. A series of numerical simulations were performed to explore the formation process of shell galaxies. In their merger scenario, the intruder dwarf galaxy is released from and disrupted immediately at a tidal point where the tidal shear from the target galaxy matches the binding force at one scale length of the dwarf galaxy. The stars stripped from the dwarf galaxy are gravitationally captured by the target galaxy. In the later stage, also named the free fall stage, the stars oscillate within the rigid potential of the target galaxy, with their energy and angular momentum conserved. These stars decelerate to near-zero radial velocity at the turning points of their orbits. The gathering of stars at the turning points induces density enhancements, which appear as ripples in the outer envelope of the target galaxy. In these studies by Quinn, the self-gravity of the dwarf galaxy and dynamical friction were neglected, resulting in constant energies of stars in the progenitor galaxy. However, observations have demonstrated that shells exist over a wide range of radii, indicating a broad range of energies for stars within the progenitor. The intruding dwarf galaxy is slowed down by dynamical friction. The stars in the progenitor lose orbital energy when they go across the target galaxy. Consequently, the energies of stars expand within a broader range \citep{Dupraz1987}. On the other hand, even in the absence of dynamical friction, the self-gravity of the dwarf galaxy significantly broadens the energy range of its stars \citep{Heisler1990}.

The studies of shell galaxies have continued to be an active area of research, including the estimation of the mass distribution and timescales after galaxy merger by measuring the number and spatial distribution of shells \citep{Quinn1984,Dupraz1986,Canalizo2007,Sanderson2013}. The gravitational potential of a target galaxy at the radii of the shells can be constrained by studying the kinematics of the shell stars \citep{MK1998,Ebrova2012,Sanderson2013,Bilek2015a,bilek2015b}. Investigations have been conducted into the formation of shell galaxies along non-radial orbital mergers \citep{Ebrova2020}, identifications of the shell structure in the Milky Way \citep{Donlon2020}, and examination of the chemical properties of shell galaxies and kinematics of shell stars \citep{Alabi2020,Fensch2020}. Research has also been carried out on the possible origins of tidal shells in the NGC 474 \citep{Bilek2022}, as well as constraints on the shapes of dark matter haloes from 2D photometric data of the extragalactic tidal streams or shells \citep{Nibauer2023}. 

The existing studies have primarily focused on examining the structure parameters of the host galaxy by analyzing the morphology and kinematics of shells. Despite the abundance of current research, there is a noticeable gap in studies focusing on the energy evolution of stars in the dwarf galaxy during the infall stage. Earlier studies have examined the changes in the orbital energy of the dwarf galaxy under the influence of dynamical friction \citep{Dupraz1987}. However, their approach involved the use of a highly compact mass model for the intruding galaxy on a parabolic orbit, thereby overlooking the role of dynamical friction during the infall stage. The orbital energy of the intruding galaxy is diminished by dynamical friction as it passes through the central region of the target galaxy in the post-infall stage. They also discussed the case of dwarf galaxies on a purely radial orbit, where dynamical friction is quite significant. In such an orbit, the specific energies of all stars are assumed to be the same as that of the intruding galaxy at all times.

Our focus is on modeling Type I and Type II shell galaxies. The orbits of stars that have been stripped from the intruder dwarf galaxy not only reflect the gravitation of the target galaxy, but also provide insightful information on the progenitor dwarf galaxy. The number and positions of observed shells at a specific time are jointly influenced by the energies of the stars, the time elapsed since the merger, and the potential of the host galaxy. We concentrate on the evolution in energies and angular momenta of individual particles within the dwarf galaxy during the infall stage, taking into account the self-gravitation of the dwarf galaxy. Contrary to the models presented in \citep{Dupraz1987}, our models of the target galaxy are compact, and the intruding galaxy is diffuse, so the latter is already disrupted while dynamical friction is still weak. By the way, if the intruding galaxy is compact, it is fully disrupted only after multiple encounters. Generally, the shells composed of stars released during the first encounter are called first-generation shells, those from the second encounter are referred to as second-generation shells, and so on \citep{bartovskova2011}. 

We lay our assumptions and semi-analytical method in Section \ref{models}, and we give the model parameters and initial conditions in Section \ref{sec:amodels}. We investigate the energy evolution of shell particles due to the infall phase using semi-analytical and compare with full N-body approaches in Section \ref{sec:EvolutionEnergies} and Section \ref{sec:Energydistribution}. 
The shell structures and differential energy distributions of intruder galaxy in the full numerical simulations are given in Section \ref{sec:radial_shellden} and Section \ref{subsec:Diff_E}. In Section \ref{mergertime} we predict the number of observable shells at any given time after the first encounter, based on the energy range of the intruder galaxy at the moment of disruption. The discussions and conclusions are presented in Section \ref{discussion}.

\section{semi-analytical models in the infall stage} \label{models}

\begin{figure}
   \centering
   \includegraphics[scale=0.32]{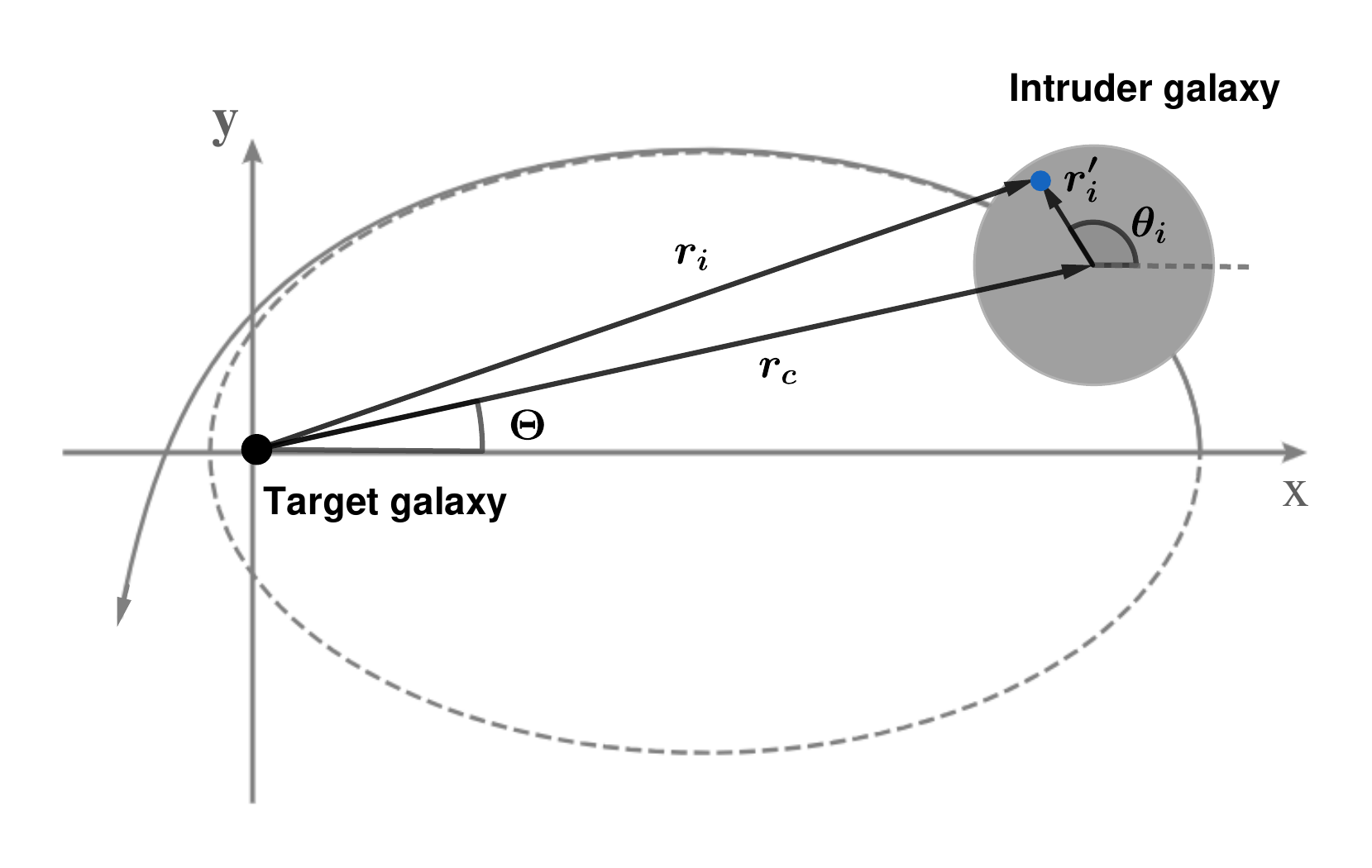} 
   \caption{shows a schematic diagram for a target or host galaxy at the origin being intruded by a rotating dwarf galaxy on a nearly-radial prograde orbit at position ($r_{c}$, $\Theta$) with velocity $\mathbf{v}_c$.  Also shown is the polar coordinates ($r_i'$,$\theta_{i}$) of the $i_{\rm th}$ star in the dwarf.  All angles are with respect to the x-axis basis $\bm{e}_{\rm x}$, pointing to the apocenter before the merger. The solid gray line represents the actual trajectory of the intruding galaxy, with arrows indicating the direction of motion, while the dashed gray line represents the Keplerian orbit.}
\label{fig1}
\end{figure}

To investigate the evolution of energies and angular momenta of the intruder dwarf galaxy particles during the merger with a massive target galaxy, we develop models according to the following assumptions: (i) that both the target galaxy and the intruder dwarf galaxy are modeled as rigid potentials, (ii) that the motion of stellar particles in the intruder galaxy follows circular orbits relative to the center of mass (CoM) of the intruder galaxy, and (iii) that the dynamical friction between the particles of the intruder galaxy and the target galaxy is neglected. The process of shell formation is divided into two stages: The ``infall stage'', defined as the period before the first encounter between the intruder and the target galaxies, and the ``free fall stage'', defined as the period after the first encounter.

We display the schematic diagram for an orbit of a galaxy merger in Fig. \ref{fig1}. The position and relative velocity vectors of the CoM of the intruder dwarf galaxy are $\rc$ and $\vc$. The position of an arbitrary stellar particle within the intruder dwarf galaxy is labeled as $\ristroke$. This particle is subjected to the gravitation $-\bm{\nabla} \Phi_1(\bm{r}_{i},t)$ from the target galaxy, the self-gravitation $-\bm{\nabla} \Phi_2({\bm{r}_{i}'},t)$ from the intruder dwarf galaxy, where ${\bm{r}_{i}'} = \ri -{\bm{r}_{c}(t)}$ is the particle's position relative to the intruder galaxy CoM. The equation of motion for the particle $i$ thus follows
\begin{equation}\label{eq1}
\dfrac{{\rm d} \bm{v}_{i}}{{\rm d}t}=-\bm{\nabla} \Phi_1({\bm{r}_{i}},t)-\bm{\nabla} \Phi_2\left[{\bm{r}_{i}} -{\bm{r}_{c}(t)},t\right].
\end{equation}
This particle has a changing energy
\begin{equation}\label{eq2}
E_{i}(t)=\dfrac{1}{2} v_i^2+\Phi_1(\bm{r}_{i},t)+\Phi_2\left[{\bm{r}_{i}} -{\bm{r}_{c}(t)},t\right].
\end{equation}
By differentiating both sides of Eq. \ref{eq2} for time and combining it with Eq. \ref{eq1}, we can derive the rate of change of energy 
\begin{equation}\label{eq3}\begin{split}
\dfrac{{\rm d}E_{i}}{{\rm d}t}
&=\bm{v}_{i}\bm{\cdot}\dfrac{{\rm d}\bm{v}_{i}}{{\rm d}t}+\dfrac{\bm{D}\Phi_1(\bm{r}_{i},t)}{\bm{D}t}+\dfrac{\bm{D}\Phi_2({\bm{r}_{i}'},t)}{\bm{D}t} \\
&=\bm{v}_{\rm c}\bm{\cdot}\left[-\bm{\nabla} \Phi_2({\bm{r}_{i}'},t)\right]+\dfrac{\partial \Phi_1}{\partial t}+\dfrac{\partial \Phi_2}{\partial t}
\end{split}\end{equation}
where we used the full derivative $\bm{D}\Phi/\bm{D}t=\partial \Phi/\partial t+\bm{v}\bm{\cdot}\bm{\nabla} \Phi$.

Eq. \ref{eq3} illustrates that self-gravitation plays a significant role in the evolution of energies of intruder particles. 
Assume neither galaxy is highly distorted before collision, the terms $\partial \Phi_1/ \partial t$ and $\partial \Phi_2/ \partial t$ on the right-hand side of the Eq. \ref{eq3} can be neglected, as if both potentials were rigid and spherical. 
Assume a specific stellar particle in the intruder galaxy moves on and stays on a circular orbit with an angular speed $ 
\omega_{i} = {v}_{i}'/{r}_{i}' =({\rm d}\Phi_2/{\rm d} r_{i}'/r_{i}')^{1/2}$ before the collision.
Under these assumptions, the variations in the energy $\Delta E_{i}$ and angular momentum $\Delta L_{i}$ of this stellar particle are dictated by Eqs. \ref{eq5} and \ref{eq6}, respectively.

\begin{equation}\label{eq5}
\begin{split}
\Delta E_{i}&\approx \int-{\bm \nabla} \Phi_2(\bm{r}_{i}')\bm{\cdot \bm{v}_{\rm c}}{\rm d}t \\
&\approx -\dfrac{{\rm d}\Phi_2}{{\rm d}r_{i}'}\int \left( v_{\rm cx}\cos\theta_{i}+v_{\rm cy}\sin \theta_{i}\right){\rm d}t,
\end{split}
\end{equation}

\begin{equation}\label{eq6}
\begin{split}
\Delta \bm{L}_{i}&\approx \int (\bm{r_{\rm c}} + \bm{r}_{i}') \times \left[-{\bm \nabla} \Phi_2(\bm{r}_{i}')\right]{\rm d}t\\
&\approx- \dfrac{{\rm d}\Phi_2}{{\rm d}r_{i}'} \int \left( x_{\rm c}\sin \theta_{i}-y_{\rm c}\cos\theta_{i} \right){\rm d}t,
\end{split}
\end{equation}
where
\begin{equation} \label{eq7}
\begin{split}
\theta_{i}=\left(\bm{e}_{\rm x},\bm{r}_{\rm i}'\right)=\omega_{i} t+\theta_{i\rm 0}.
\end{split}
\end{equation}
In Eq. \ref{eq7}, the initial phase angle $\theta_{i\rm 0}$ of the particle $i$ with a fixed circular frequency $\omega_{i}$ is measured from the x-axis basis $\bm{e}_{\rm x}$.

To derive the changes of energies and angular momenta for the particles from the intruder galaxy, we need to know about the temporal evolving position, $(x_{\rm c}, ~y_{\rm c}, ~z_{\rm c})$, and velocity, $(v_{\rm cx}, ~v_{\rm cy}, ~v_{\rm cz})$, of the CoM of the intruder galaxy. Disregarding the dynamical friction between the intruder and target galaxies, the CoM velocity of the intruder galaxy can be calculated as follows
\begin{itemize}
\item (i) When the intruder galaxy is falling along a radial orbit, the orbital position $( x_{\rm c}(t),0,0)$  can be calculated directly from the energy conservation, 
\bey\label{radialE}
\int_{x_{\rm c}(0)}^{x_{\rm c}(t)} \frac{{\rm d} x_{\rm c}}{\sqrt{2\left[E_{\rm c}-\Phi_1(x_{\rm c}))\right]}}=t.
\eey
\item (ii) When the intruder galaxy is orbiting the target galaxy of mass $M_1$ along a nearly-radial orbit of a specific orbital angular momentum $L_{\rm c}$ and energy $E_{\rm c}$, the orbit is approximated as an ellipse, 
\bey\label{ellipse}
r_{\rm c}(t)&=&\frac{L_{\rm c}^2}{GM_1 \cdot \left[1-e\cos \Theta(t)\right]}, \\
e&=&\sqrt{1+\frac{2E_{\rm c}L_{\rm c}^2}{G^2M_1^2}},\nonumber
\eey
where $e$ is the ellipticity of this orbit, and the time $t$ is derived from the conservation of angular momentum,
\beq
t = L_{\rm c}^{-1} \int_0^{\Theta(t)} r_{\rm c}^2 d\Theta.
\eeq
\end{itemize}

Thus, for given initial orbital parameters of the intruder galaxy, Eqs. \ref{eq5} and \ref{eq6} imply that the evolution of the energy and angular momentum of an arbitrary particle can be predicted by given initial values of $r_{i\rm 0}'$ and $\theta_{i\rm 0}$ with the following equation, 
\beq\label{eqenergy} E_{\rm i}=E_{i\rm 0} + \Delta E_{i}. \eeq
The initial energy of the particle is
\beq\label{eini} E_{i\rm 0}\approx \Phi_1(r_{\rm c0})+\Phi_2(r_{i\rm 0}')+\frac{1}{2}\left[v_{\rm c0}^2+ v_{i\rm 0}'^2+2v_{\rm c0}v_{i\rm 0}'\sin(\theta_{i\rm 0})\right]. \eeq
We approximate $\Phi_1(r_{i}) \approx \Phi_1 (r_{\rm c0})$, since $r_{i}\approx r_{\rm c0}$ and the initial kinetic energy of the particle depends on $\theta_{i\rm 0}$. Therefore, when the model parameters are determined, $E_{i}$ can be determined by ($r_{i\rm 0}'$, $\theta_{i\rm 0}$). We can also relate $E_{i}$ to ($r_{i}'$, $\theta_{i}$) using Eq. \ref{eq7}. The coordinates ($r_{i}'$, $\theta_{i}$) denote the positions of stars within the intruder at its pericenter.
\begin{table*}
\centering
\caption[]{lists the model parameters of ($M_1$, $r_{s1}$, $N_1$) for the mass, the scale length, and the live particle number for the target galaxy, and ($M_2$, $r_{s2}$, $N_2$) for those of the intruder galaxy, as well as the intruder's velocity anisotropy $\beta$, initial position and velocity, and orbital angular momentum $L_c$. }\begin{tabular}{ccccccccccccc}
\hline
Model & $\rho_{1}$ & $M_{1}$ & $r_{\rm s1}$ & $N_{1}$ & $M_{2}$ &$r_{\rm s2}$ & $N_{2}$ & $\beta$ &  $(x_0,~y_0,~z_0)$& $(v_{\rm x0},~v_{\rm y0},~v_{\rm z0})$  & $L_{\rm c}$\\
 &  & $(10^{10}\msun)$ & $(\kpc)$ & $(10^5)$ &   $(10^{10}\msun) $ & $(\kpc)$ & $(10^5)$ & & $(\kpc)$ & $(\kms)$ & $(\kpc \kms)$\\
 \hline
PAR & Plummer & 200 & 12.88 & 10  & 2 & 5.55 & 2 & $-\infty$ & (150,~0,~0) &(-189,0,0) & 0\\
PIR & Plummer & 200 & 12.88 & 10  & 2 & 5.55 & 2 & $0$ &(150,~0,~0) &(-189,0,0)&0 \\
HAR & Hernquist & 200 & 12.88& 10  & 2 & 5.55 & 2 & $-\infty$ & (150,~0,~0) &(-177,0,0) &0\\
1PANR & Plummer & 200 & 12.88 & 10  & 2 & 5.55 & 2 & $-\infty$ & (150,~15,~0) &(-239,0,0) & 3583 \\
2PANR & Plummer & 200 & 12.88 & 10  & 2 & 5.55 & 2 & $-\infty$ & (150,~30,~0) &(-237,0,0) & 7113 \\
3PANR & Plummer & 200 & 12.88 & 10  & 2 & 5.55 & 2 & $-\infty$ & (150,~0,~0) &(0,50,0) &7500 \\
\hline
\end{tabular}\label{orbit}
\end{table*}

\begin{figure*}
\centering
\includegraphics[scale=0.69]{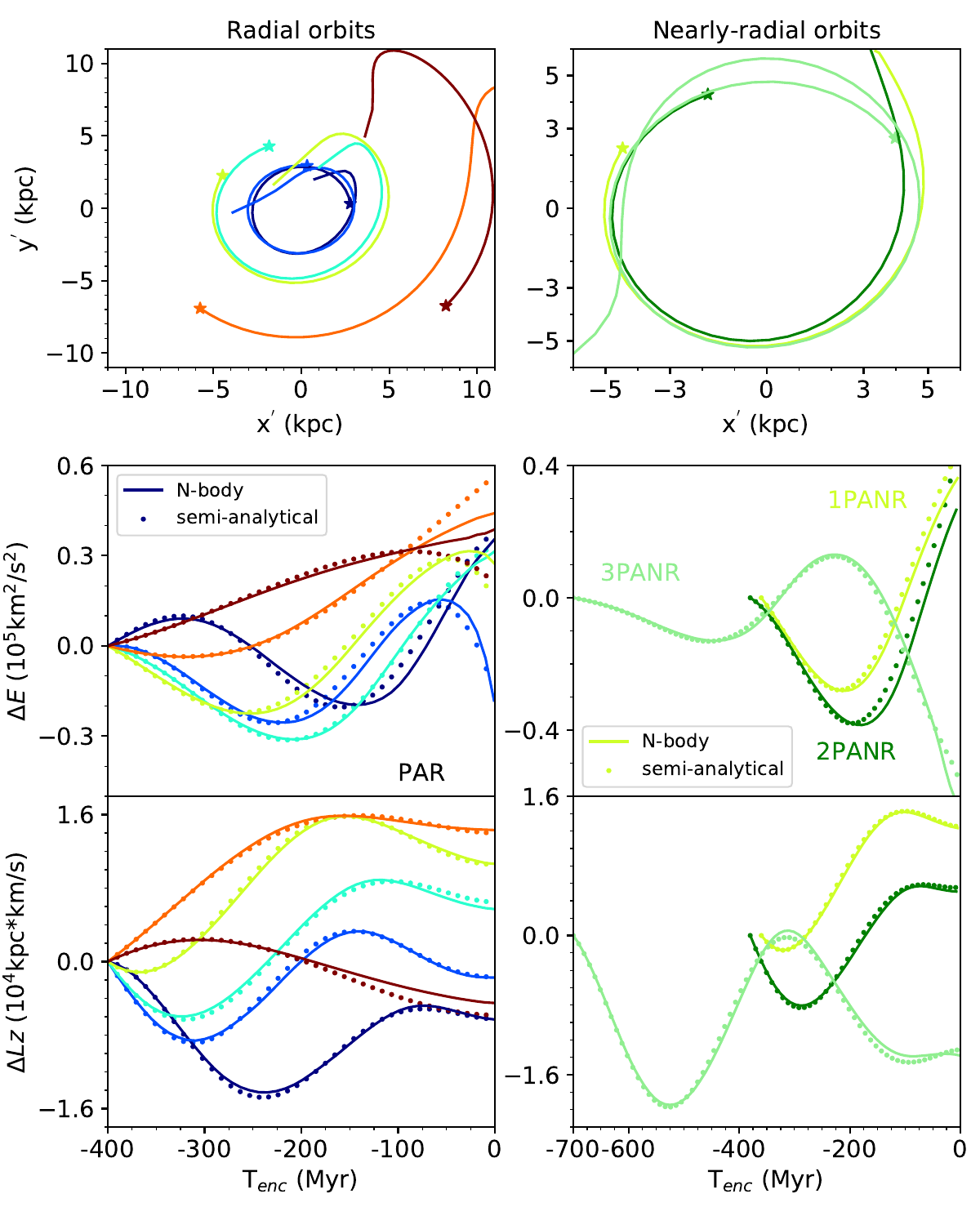}
\caption{shows colored orbits of particles of various initial conditions (top panel) from the intruder galaxy, and their energy $E$ (middle) and angular momentum $L_z$ (bottom) in the radial collision Model PAR (left panels), and near-radial collision Models 1PANR, 2PANR and 3PANR (right panels) as functions of the lookback time $T_{\rm enc}$. The evolution in N-body runs (solid lines) is well-matched by our analytical predictions (colored dots).}
\label{fig:rcsall}
\end{figure*}

\section{Comparison of semi-analytical model and numerical simulations}
\label{sec:aEvolutionEnergies}
\subsection{Models}
\label{sec:amodels}

To investigate the dynamics of shells generated through mergers and verify our analytical model, we perform a series of N-body simulations. Both the intruder and target galaxies are composed of a large number of live particles. The N-body simulations can yield outcomes that bear a close resemblance to actual galaxies, comprehensively incorporating various physical factors. These include the self-gravity of the dwarf galaxy and the effects of dynamical friction. 

\subsubsection{Density and velocity profiles}
Considering that shells typically appear around elliptical galaxies \citep{Malin1983,duc2015,bilek2016}, we simulate the formation of shell galaxies within a spherically symmetric potential of the target galaxy. To validate the applicability of our simplified model, we employed two types of spherically symmetric potential models for the target galaxy. These models can be universally described by the following formula \citep{zhao1996},

\begin{equation}
\label{eq9}
\begin{split}
\Phi\left(r \right) = -\dfrac{GM_1}{r_{\rm s1}} \left[1+\left(\dfrac{r}{r_{\rm s1}}\right)^{1/\alpha}\right]^{-\alpha},
\end{split}
\end{equation}
\begin{equation}
\label{eq8}
\begin{split}
\rho\left(r \right) = \rho_{0}\left(\dfrac{r}{r_{\rm s1}}\right)^{-\gamma} \left[1+\left(\dfrac{r}{r_{\rm s1}}\right)^{1/ \alpha}\right]^{(\gamma - \xi)\alpha}, \\~ [\gamma, \xi, \rho_0] = [2 - \frac{1}{\alpha},  3 + \frac{1}{\alpha}, \left(1+\frac{1}{\alpha}\right) \frac{M_1}{4\pi r_{\rm s1}^3}],
\end{split}
\end{equation}
where $\alpha$, $\xi$ and $\gamma$ describe density double-power-law slopes. The values of $(\alpha, ~\xi, ~\gamma )$ taking $(1/2,~5,~0)$ represent a Plummer's profile \citep{Plummer1911} with $\rho_{0}=3M_1/(4\pi r_{\rm s1}^3)$, while taking $(1,~4,~1)$ stand for a Hernquist's model \citep{Hernquist1990} with $\rho_0$ being $M_1/(2\pi r_{\rm s1}^3)$. In the above formula, $M_1$ and $r_{\rm s1}$ are the total mass and the scale length of the target galaxy, respectively. In contrast, the Hernquist model exhibits a pronounced central peak in both density and potential profiles, while the Plummer model demonstrates a flat core in the center in both density and potential profiles. 

The anisotropy of velocity dispersion of the particles in a galaxy is defined as
\beq \beta \equiv 1-\left(\overline{v_{\rm \theta}^2} + \overline{v_{\rm \phi}^2}\right)/ \left(2\overline{v_{\rm r}^2}\right),\eeq
where $\overline{v_{\rm r}^2}$ is the radial velocity dispersion, and $\overline{v_{\rm \theta}^2}$ and $\overline{v_{\rm \phi}^2}$ are the polar and azimuthal velocity dispersions. The target galaxy exhibits an isotropic velocity dispersion, characterized by a $\beta$ value of 0. 

For the intruder dwarf galaxy, we consider two anisotropic models for the velocity distributions. One of them is a tangentially anisotropic model with an anisotropy parameter of $\beta=-\infty$. In this model, all particles follow circular orbits, with their speeds determined by
\beq{v_{i}'}^2=\frac{GM_2{r_{i}'}^{1/\alpha}}{ \left({r_{i}'}^{1/\alpha}+r_{\rm s2}^{1/\alpha}\right)^{1+\alpha}}. \eeq
The velocity directions are assigned randomly. Here $M_2$ and $r_{\rm s2}$ represent the total mass and the scale length of the intruder galaxy, respectively. The $\alpha=1/2$ and $\alpha=1$ mass models are adopted with a tangentially anisotropic velocity dispersion profile. 
The other velocity model for the intruder galaxy is isotropic ($\beta = 0$). We only apply this to the Plummer mass model, where 
\beq\overline{v_{\rm r}^2}=\overline{v_{\rm \theta}^2}=\overline{v_{\rm \phi}^2}= \frac{GM_2 }{ \left(6\sqrt{{r_{i}'}^2+r_{\rm s2}^2}\right)}.\eeq

\subsubsection{N-body ICs and numerical tools}

To generate the N-body live particle initial conditions (ICs), we use a publicly available code {\it McLuster} \citep{kupper2011} to generate the Plummer spheres for both target and intruder galaxies. The ICs for the target galaxy with a Hernquist profile are generated by a {\it Disk Initial Conditions Environment} code \citep[{\it DICE},][]{Perret2016}. {\it DICE} determines the superpositions of particles by applying a Metropolis-Hasting Monte-Carlo Markov Chain algorithm utilizing gravitational acceleration. The numbers of particles for the target and the intruder galaxies are $10^6$ and $2.0 \times 10^5$ respectively. The density profiles of the target galaxy models adhere to two forms and are summarised in Table \ref{orbit}. The first (for Models PAR, PIR, 1PANR, 2PANR, 3PANR) is a Plummer profile, with a total mass of $ M_1 = 2.0\times 10^{12} \Msun$ and a scale length of $r_{\rm s1} = 12.88 ~\kpc$. The second form (for Model HAR) is a Hernquist profile, which maintains the same mass of $M_1$ and the same scale length of $r_{\rm s1}$. The total mass of the intruder galaxy is $M_2 = 2.0\times 10^{10} \Msun$, with a scale length of $r_{\rm s2} = 5.55$ kpc. At the initial state, both the target and intruder galaxies are in virial equilibrium. We will perform eight N-body simulations for an intruder galaxy moving on both radial (Models PAR, PIR and HAR) and nearly-radial (Models 1PANR, 2PANR and 3PANR) orbits. The initial positions for the intruder galaxy are listed in Table \ref{orbit}, together with the initial relative net velocities. Since the stars of the shells are originally from the intruder dwarf galaxy, we need to differentiate the stellar and dark matter halo components from the single Plummer sphere. The stellar particles of the intruder galaxy are characterized following the definitions adopted by \citet{Bret2017} and \citet{Hendel2015}, specifically as the particles that possess the top $10\%$ of the tightest binding energy. 
 
We perform N-body simulations by using a publicly available Adaptive Mesh Refinement (AMR) code, {\it RAMSES} \citep{Teyssier2002}. {\it RAMSES} is a code that solves the Poisson's equation via the Particle Mesh (PM) method on AMR grids \citet{Guillet2011}. In this study, the N-body simulations are performed within a cubic box of $(5000 ~\kpc)^3$ using a static coarse grid segmented into  $128^3$ cells. Additional grid refinement is triggered if more than five particles are in a cell. The cell is split into eight sub-cells. The maximal level of refinement is $l_{\rm max}=24$ to reach the maximal resolution of approximately $0.3~ \pc$. The actual maximal level of refinement in the simulations is $l_{\rm max}^{\rm actl}=16$, corresponding to an actual maximal spatial resolution of $76~\pc$.

\subsection{\texorpdfstring{Inferring $\Delta E_i$ and $\Delta Lz_i$ of particles in the infall stage and comparing with numerical simulations}{Inferring changes of energy and angular momentum of particles in the infall stage and comparing with numerical simulations}}
\label{sec:EvolutionEnergies}

To validate the semi-analytical model and better understand the evolution of energy and angular momentum, i.e., $\Delta E$ and $\Delta L_{z}$, for particles in the intruder galaxy, we carry out an in-depth analysis of the radial (sharing the same parameters for mass profile and orbit as Model PAR) and nearly-radial merger models (possessing the same parameters as Models 1PANR, 2PANR and 3PANR). We analyze $\Delta E$ and $\Delta L_{z}$ of stars on the infalling orbits of the intruder galaxy up until the first encounter using our semi-analytic method. We compare the above results with the outcomes from N-body simulations.

Model PAR is an N-body simulation of an intruder galaxy merging with a target galaxy along a radial orbit, characterized by zero orbital angular momentum. In the Model PAR N-body simulation, all particles in the intruder galaxy move in circular orbits around the galaxy center ($\beta=-\infty$, see Table \ref{orbit}). We randomly choose three groups of particles, each located at different radii aligned with the orbital plane of the intruder galaxy in the infall stage. We define this plane as the merger plane. Each group consists of two particles with different initial phase angles (i.e., the azimuthal angles of particles on the merger plane, denoted as $\theta_{i\rm 0}$ for the $i_{\rm th}$ particle as aforementioned), rotating counterclockwise relative to the center of the intruder galaxy. We define $T_{\rm enc}=0 \Myr$ as the time of the first encounter between the two galaxies, after which is the free fall time of stars originated from the intruder galaxy. The top left panel of Fig. \ref{fig:rcsall} shows the trajectory of particle motion around the center of the intruder galaxy, with the initial positions of the particles indicated by asterisks.
As the intruder galaxy moves toward the center of the target galaxy, these particles experience ongoing shifts in energy and angular momentum (middle/bottom panel). In this stage, the particles complete at least half a circular period around the center of the intruder galaxy.

Let's focus on the navy blue particle situated at a radius of 3kpc. This particle, with an initial phase angle near $0^\circ$, moves counterclockwise in the first, second, third, and fourth quadrants. According to Eqs. \ref{eq5} and \ref{eq6}, we can deduce that the sequence of the rate of energy changes, $dE/dt$, for particles is respectively $> 0$, $< 0$, $< 0$ and $ > 0$ within the four quadrants. Meanwhile, the sequence of the rate of angular momentum changes, $dL_{\rm z}/dt$, is  $<0$, $< 0$, $ > 0$, and $ > 0$. Fig. \ref{fig:rcsall} confirms that the evolution of energies and angular momenta of these particles indeed follow this pattern. By examining the dashed and solid curves, it is evident that the analytically predicted $\Delta E$ and $\Delta L_{z}$ align closely with the results from the N-body simulation of Model PAR. However, a minor discrepancy is observed in the predicted results after $T_{\rm enc}=-100~\Myr$ compared to the simulation outcomes. In the N-body simulation, when the intruder galaxy comes within $40~\kpc$ of the target galaxy, deformation and disruption are inevitable. This causes the particles of the intruder galaxy to stray from their original orbits, resulting in discrepancies in the analytically predicted outcomes. The analysis of these particles also applies to other particles. For particles at larger initial radii, the simulation results diverge from the analytically predicted behavior sooner, as these outer particles deviate from their original orbits earlier.

In near-radial orbit Models 1PANR and 2PANR, the intruder galaxies share identical initial orbital energies on orbits with different angular momenta. Model 3PANR starts with a much lower kinetic energy. All these models are disrupted by the target galaxy when approaching the pericenters of 5-10 kpc. The right panel of Fig. \ref{fig:rcsall} shows $\Delta E$ and $\Delta L_{z}$ of three shell particles {which are} randomly selected at $5~\kpc$ from the intruder center. The initial phase angles of the selected particles in Models 1PANR, 2PANR and 3PANR are different. We find that the analytic predictions agree well with the results derived from N-body simulations. Moreover, the particle selected from Model 3PANR exhibits a notably larger count of rotations around the center of the intruder galaxy, compared to particles from Models 1PANR and 2PANR, in the upper right panel of Fig. \ref{fig:rcsall}. This is attributed to the longer infall timescale of Model 3PANR, in which the intruder galaxy is launched from the apocenter of a nearly-radial orbit.

In conclusion, $\Delta E$ and $\Delta L_{z}$ of these particles exhibit periodic behaviors when the intruder galaxy is moving along a radial orbit. The evolution of energy of an $i_{\rm th}$ particle, $\Delta E_{i}$, is related to the value of $\cos (\theta_{i})$, where $\theta_{i}$ is the angle between the vector $\bm{r}_{i}'$ and the positive x-axis. When $\cos (\theta_{i}) > 0$, there is a positive rate of change in the particle energy, whereas the rate of change becomes negative when $\cos (\theta_{i}) < 0$. The rate of change in angular momentum mirrors the evolution of energy but with a phase lag of $\dfrac{\pi}{2}$.
When the intruder galaxy is moving along a nearly-radial orbit, $\Delta E$ and $\Delta L_{z}$ of all particles can still be described by Eqs. \ref{eq5} and \ref{eq6}. Due to the non-rigid nature of the intruder galaxy, $\theta_{i}$ continuously varies as the intruder galaxy moves towards the target galaxy. Particles originating from different positions, represented as $\bm{r}_{i\rm 0}'$, have distinct intrinsic periods. Consequently, at any given instant after the first encounter between the two galaxies, the evolution of net energies and net angular momenta of particles is determined by both the initial radius ${r}_{i\rm 0}'$ and the initial phase angle $\theta_{i\rm 0}$.

\subsection{Inferring energy distribution of particles in the infall stage}
\label{sec:Energydistribution}

\begin{figure*}
\centering
\includegraphics[scale=0.55]{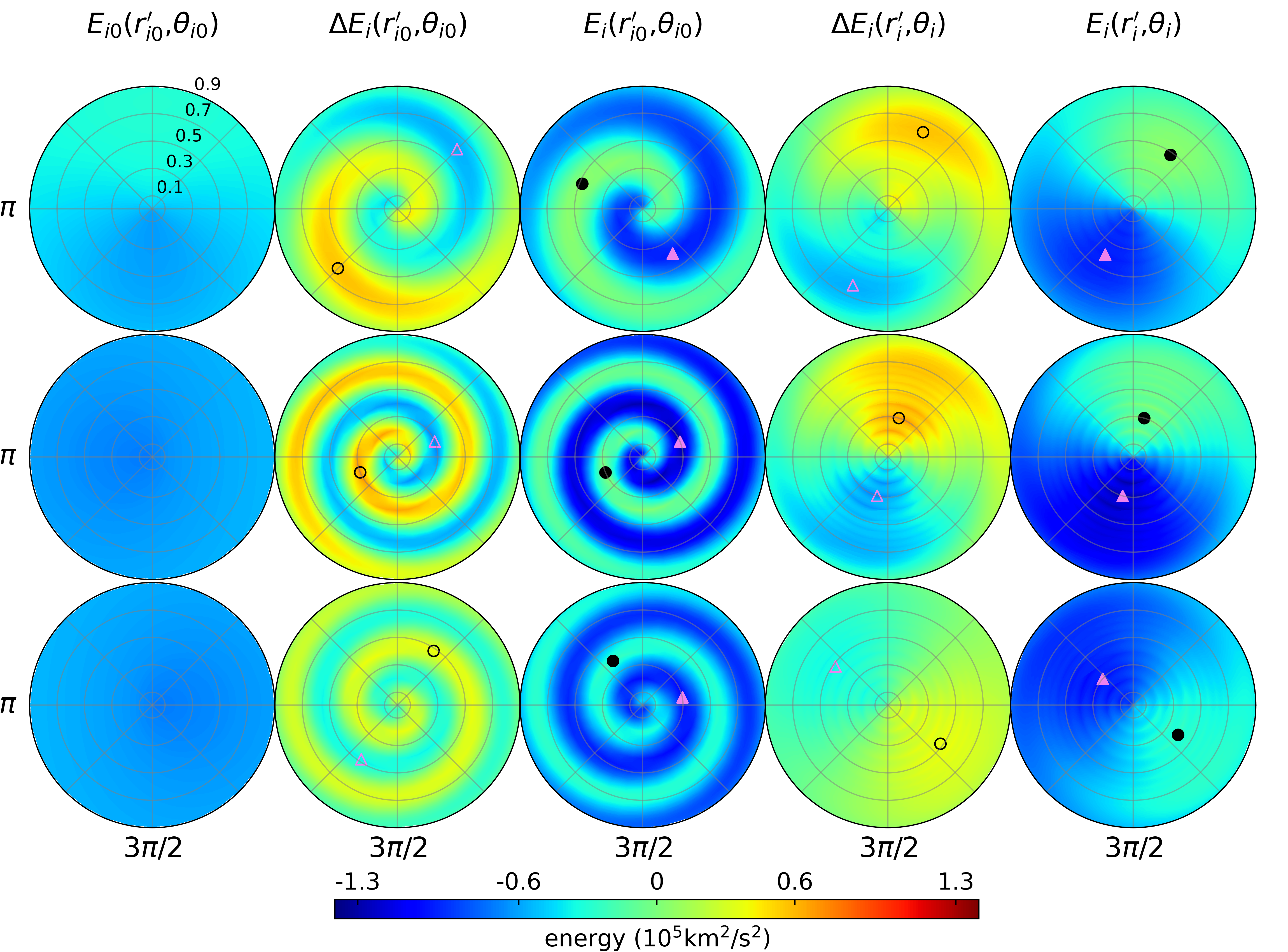} 
\caption{This figure displays the energy distribution of particles as a function of particle positions ($r_i'$,$\theta_i$) in the orbital plane of the intruder galaxy on a purely-radial orbit (1st row, with the same parameters as Model PAR), and nearly-radial prograde/retrograde orbits (2nd/3rd rows, using the same parameters as Model 3PANR) at the moment of disruption. The energy distributions are shown in polar coordinate systems with the enclosed mass fraction, $f=\frac{M_2(r_{i}')}{M_2}$ (where $M_2(r_{i}')=M_2\left(\frac{r_{i}'}{r_{s2}}\right)^3/\left[1+\left(\frac{r_{i}'}{r_{s2}}\right)^2\right]^{1.5}$)
, being the radial coordinates, and the phase angle $\theta_i$ being the azimuthal coordinates. Column 2 depicts the large change of energy distribution from the (almost uniform) initial distribution (Column 1), and final distribution (Column 3), all expressed as functions of the initial positions of particles on the orbital plane ($r_{i0}'$,$\theta_{i0}$). The same change of energy distribution from the initial (Column 1) to the final (Column 5) is depicted again in Column 4, but both are expressed as functions of the positions ($r_i'$,$\theta_i$) at the epoch of dwarf disruption. Particles that gain the most energy (hollow black circles) and lose the most energy (hollow purple triangles) are marked, as well as particles with the maximum energy $E_{max}$ (solid black circles) and the minimum energy $E_{min}$ (solid purple triangles). 
In each panel, the five circles respectively enclose 10\%, 30\%, 50\%, 70\%, and 90\% of the total mass of the intruder galaxy.}
\label{fig:CCWCW} 
\end{figure*}

Particles in the intruder galaxy that move perpendicular to the merger plane theoretically exhibit zero energy variation. This is owing to the perpendicularity between the self-gravitation $-\bm{\nabla} \Phi_2({\bm{r}_{i}'})$ and ${\bm v}_{\rm c}$. 

Our main focus of analysis is on particles that move parallel to the merger plane. These particles can be further categorized into prograde particles and retrograde particles. In our model, prograde particles refer to particles rotating counterclockwise relative to the CoM of the intruder galaxy. The angular momentum of the prograde particles is aligned in the same direction as the CoM angular momentum. Retrograde particles, on the other hand, refer to particles rotating clockwise relative to the CoM of the intruder galaxy, with their angular momentum in the opposite direction to that of the CoM. The definitions of prograde and retrograde particles remain consistent with the existing studies \citep[e.g.,][]{toomre1972,read2006,duc2013}.

According to the semi-analytical model, we have predicted the energy distributions of prograde and retrograde particles. In the case of a purely radial motion of (the CoM of) the intruder dwarf galaxy sharing the same parameters for density profile and orbit as Model PAR, the energy distributions of prograde and retrograde particles exhibit mirror symmetry. Therefore, we only display the energy distribution of prograde particles as shown in the first, i.e., uppermost row of Fig. \ref{fig:CCWCW}. However, in the case of nearly-radial motion of the intruder dwarf galaxy, using the same parameters as Model 3PANR, there is an excursion between the energy distributions of prograde and retrograde particles, as depicted in the second and third rows of Fig. \ref{fig:CCWCW}, respectively. We note that the energy changes of particles originating from different initial positions are different, similar to their final energy. Here, ``final energy" refers to the energy of particles when the intruder galaxy reaches the pericenter of its orbit. In our model, the intruder dwarf galaxy is instantaneously disrupted at the pericenter, such that the final energy can be interpreted as the particle energy at the moment of release. 

In the case of purely radial motion, the intruder dwarf galaxy displays a smaller initial energy dispersion, while the range of final state energy expands significantly, approximately three times the initial energy range. In the case of nearly-radial motion, we have found that the energy changes of retrograde particles are less pronounced compared to those of prograde particles. This difference is attributed to the resonance effect between the angular frequency of the prograde particles and the orbital frequency of the dwarf galaxy. For prograde particles, the range of final state energy is eight times broader than that of the initial energy range, while for retrograde particles, this ratio is only five times.

In the case of nearly-radial motion, the intruder galaxy takes a longer time to reach the pericenter, resulting in a longer tadpole tail. The same color in the second and fourth columns respectively indicates the initial positions and the positions at the epoch of dwarf disruption, and the color represents the net change in the energy of the particles. The fourth column shows that particles released from the ``front seats" (referring to the second and third quadrants) of the intruder dwarf galaxy lose energy, while those from the ``back seats'' (the first and fourth quadrants) gain energy. This pattern aligns with the energy evolution behavior of particles shown in Fig. \ref{fig:rcsall}. We use hollow black circles and purple triangles to mark the particles that gain the most energy and lose the most energy, respectively. We find that the particles that gain the most energy and lose the most energy end up in the back positions and front positions, respectively. The particles with the maximum energy $E_{\rm max}$ (solid black circles) and the minimum energy $E_{\rm min}$ (solid purple triangles) demonstrate similar behavior. Ripples appear in the fourth column in models moving along nearly-radial prograde/retrograde orbits. The ripples might be a result of orbital resonances. The study of orbital resonances has far exceeded the scope of this paper and we shall not delve further into it.

During the free fall stage, the minimum and maximum energies, $E_{\rm min}$ and $E_{\rm max}$, are crucial quantities for predicting the number of 
shell structures at a given time. The minimum energy particles will contribute to the formation of the innermost shell layers, while the higher energy particles determine the formation of the outer shell structures. The maximum particle energy exceeds zero, there is no concern about a lack of particles participating in the formation of the outermost shell structure. If we ignore the energy evolution, then in the free fall stage, we would see fewer shell structures, with a reduction in both the inner and outer shells. This is because the energy evolution expands the energy range, affecting both the low and high-energy ends.

\begin{figure*}
\centering
\includegraphics[scale=1.3]{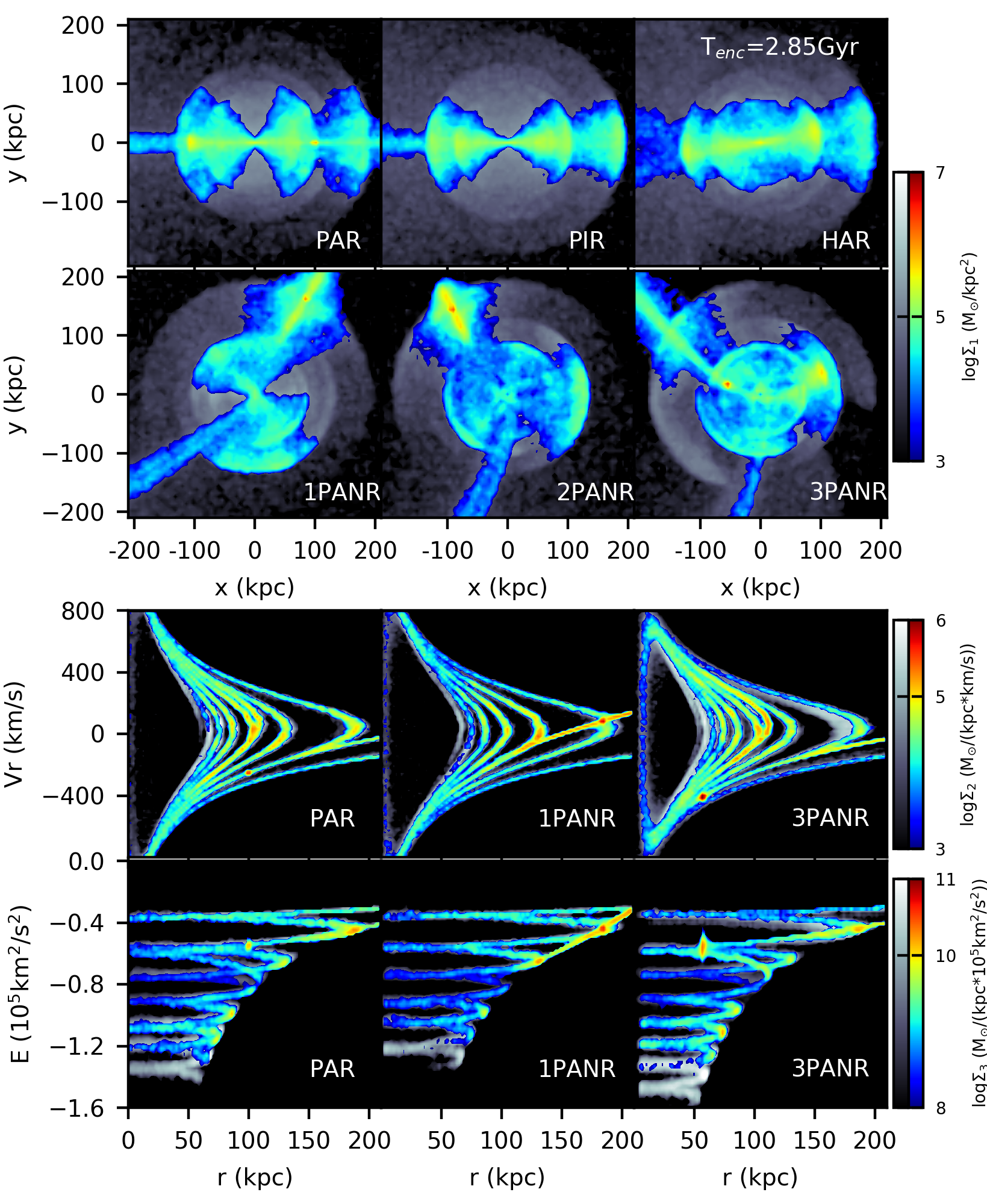}
\caption{This figure displays the projected densities (upper and second row of panels) of the shell structures formed at $T_{\rm enc}=2.85~\Gyr$ in Models PAR, PIR, HAR, 1PANR, 2PANR and 3PANR. The phase space densities (panels in the third row) and the energy densities (bottom panels) of the shell structures formed at $T_{\rm enc}=2.85~\Gyr$ in Models PAR  (panels in the left column), 1PANR (panels in the middle column) and 3PANR (panels in the right column) in the N-body simulations. The rainbow and grey color tables illustrate these densities of the stellar and the overall particles of the intruder galaxy, respectively.}
\label{fig:r_denvrE} 
\end{figure*}

In conclusion, during the process of the intruder galaxy approaching the target galaxy, the energy of particles in the intruder galaxy undergoes remarkable changes. Regardless of whether the particles are prograde or retrograde, those that are ultimately released from the ``front seats" of the intruder dwarf galaxy lose energy, while particles that are released from the ``back seats" gain energy. We find that the energy changes induced by the self-gravity of the progenitor are remarkable. The range of energy for stellar particles in the progenitor is a few times broader than that in Quinn's (1984) model. To obtain the same range of energy for stellar particles in shells, Quinn's original model needs a more massive intruding galaxy.

\section{Full N-body simulations}
\label{sec:shellNS}
\subsection{Shell structures}
\label{sec:radial_shellden}

After the first encounter, shell structures gradually form. As time progresses, both the quantity of shells and their respective radii continue to grow. The first and second rows of Fig. \ref{fig:r_denvrE} present the projected densities of the shell structures. These structures formed at $T_{\rm enc} = 2.85~ \Gyr$ in Models PAR, PIR, HAR, 1PANR, 2PANR and 3PANR. Since the shells are remnants of the merged intruder galaxy, only the particles from the intruder galaxy are shown.

In Models PAR, PIR and HAR where the intruder galaxies move along radial orbits, the shells are axially symmetric. They are distributed on both sides of the target galaxy, as demonstrated in the uppermost panels of Fig. \ref{fig:r_denvrE}). At $T_{\rm enc} = 2.85~\Gyr$, five noticeable shells appear. The furthest shell extends out to $200 ~\kpc$, exceeding the initial distance of the intruder galaxy. The numbers and positions of the shells formed in Models PAR and PIR are almost indistinguishable, except for a small structure appearing after the first encounter in Model PAR. The small structure that remains in existence is the remnant of the intruder galaxy that has not yet been disrupted. Model PAR is the most stable and hardest to destroy. This is because the intruder galaxy is rotationally supported. The energy in such a galaxy is much lower at the center than in a galaxy supported by velocity dispersion. This is why a remnant is found in Model PAR. For the comparison of Models PAR and HAR, no significant differences are observed in the number and positions of the shells. However, the intruder galaxy has been completely disrupted in Model HAR. This could potentially be attributed to the presence of a sharp central density peak in the Hernquist profile of the target galaxy.

The shell structures formed through mergers on nearly-radial (Models 1PANR, 2PANR and 3PANR) orbits may appear in a wider range of angular directions surrounding the target galaxy center (the second row of Fig. \ref{fig:r_denvrE}). The projected density maps agree with the appearance of the observed Type II shell galaxies. The formation of shells in Model 1PANR exhibits a narrower range of angular directions compared to Models 2PANR and 3PANR. This could potentially be attributed to the smaller initial orbital angular momentum present in Model 1PANR. In Models 1PANR and 2PANR, five first-generation stellar shells are formed. In contrast, Model 3PANR features an additional innermost first-generation stellar shell at the same timescale, which has lower energy. Note that the difference lies in the innermost shell, located around $60\kpc$ to the host center, which is difficult to identify from the projected density profiles. It is easier to be identified from the phase-space density distribution (third row of Fig. \ref{fig:r_denvrE}) and the energy space densities (fourth row of Fig. \ref{fig:r_denvrE}).

From the density maps, we observe that the stellar shells align in quantity and position with the shells composed of all particles. These stellar particles, initially the lowest energy particles from the intruder galaxy, contribute to both low and high-energy shells. This indicates a significant energy shift in these particles during the galaxy merger. This is consistent with the findings in Sections \ref{sec:EvolutionEnergies} and \ref{sec:Energydistribution}.

In addition to the density maps, we also examine the shell distribution in phase spaces of Models PAR, 1PANR and 3PANR. As shown in the third row of panels of Fig. \ref{fig:r_denvrE}, the shell structures are more easily distinguishable when viewed in phase space. 
In Models PAR and 3PANR, the third shell seems to partially split into two separate shells, one of which is likely a second-generation shell.

In the energy space diagrams, the shells display a distinct layering pattern after the first encounter. They also exhibit turning points that mirror those in the phase space, aligning with the shell positions shown in Fig. \ref{fig:r_denvrE}. Upon the formation of the shells, the stellar particles reach a minimum energy of approximately $-1.3\times10^5 ~\kmssq$. All of the shells are found to be assembled within the energy range of $-1.3\times 10^5~ \kmssq$ and $0~ \kmssq$. It is worth noting that at a large radius of $ r \gg 200~\kpc$, a branch of particles with their energy exceeding zero. The escaping stellar particles are not presented in the bottom panels of Fig. \ref{fig:r_denvrE}, since they are beyond $200~\kpc$. The existence of escaping particles is consistent with the analysis in Section \ref{sec:Energydistribution}. 

On top of that, in the N-body simulations, the stellar shells exhibit an overlap with the shells constituted by all particles. This overlap is discernible in the context of projected, phase space, and energy densities. However, it's noteworthy that these densities of the stellar components are significantly more concentrated within the ICs.

\subsection{Differential energy distribution}
\label{subsec:Diff_E}

\begin{figure}
\centering
\includegraphics[scale=0.7]{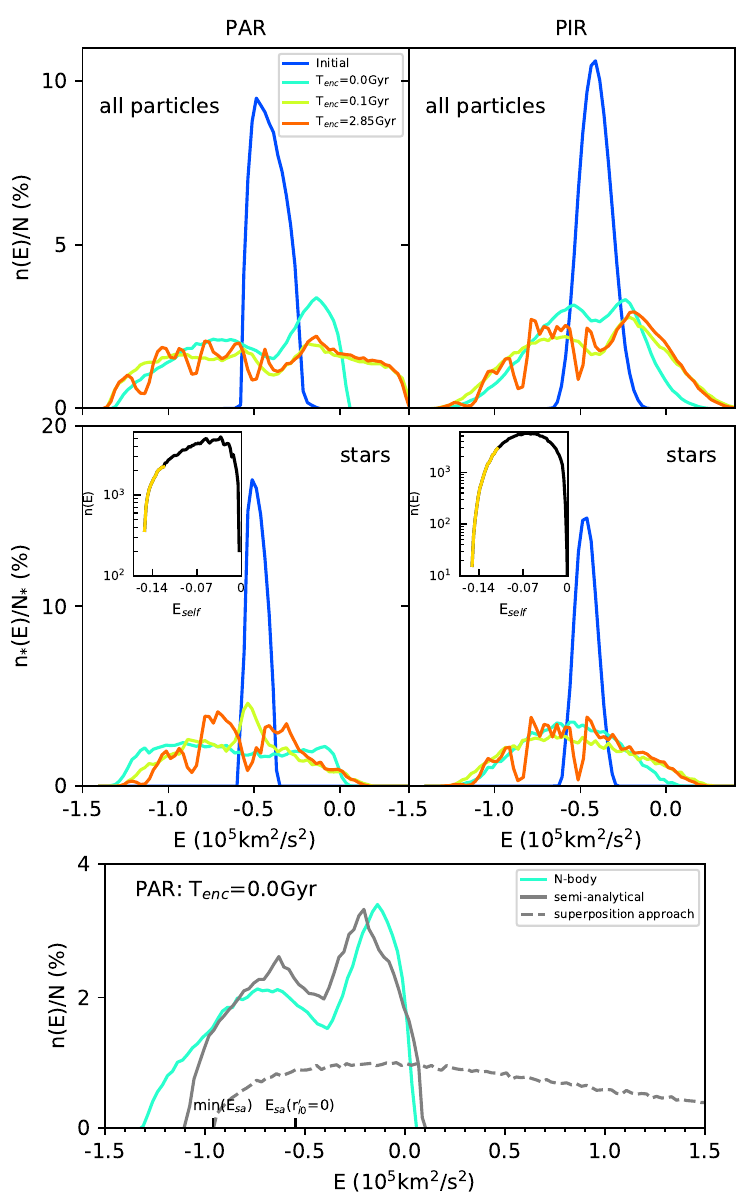} 
\caption{The differential energy distributions $n(E)/N$ of overall (top) and stellar particles (middle) of the intruder galaxy are shown at four snapshots of simulations (at the beginning: blue curve, at $T_{\rm enc}=0~\Gyr$: cyan, at $T_{\rm enc}= 0.1 ~\Gyr$: green, at $T_{\rm enc}=2.85~\Gyr$: orange). These panels correspond to the two N-body models: Models PAR and PIR from the left to right panels. In the two smaller panels appearing in the lower panels, the black line shows the differential energy distribution of all particles $n(E)$ relative to the dwarf galaxy itself at the beginning, while the yellow line represents the stellar particles. The bottom panel shows the comparison of differential energy among the N-body simulation of Model PAR (cyan solid line), the semi-analytical method (gray solid line) and the superposition approach,  with the latter two methods having the same parameters for the density profile and orbit as Model PAR, at $T_{\rm enc}=0~\Gyr$. The lowest energy and the energy of the central particle calculated by the superposition approach are marked as $min(E_{\rm sa})$ and $E_{\rm sa}(r_{i0}'=0)$, respectively. The energy units of all panels are consistent.}
\label{fig:NEuse}
\end{figure}

The differential energy distributions of all particles n(E) within the intruder galaxy, are defined by the relation
\bey
\int_{E_{\rm min}}^{E_{\rm max}} n(E) dE=N.
\eey
The range of integration is from the minimum to the maximum energies attained by the particles. $N$ is the total number of particles in the intruder galaxy.

We study the differential energy distributions of the intruder galaxy at four different time snapshots: the initial state, the moment of first encounter ($T_{\rm enc}=0$), $T_{\rm enc}=0.1 ~\Gyr$, and $T_{\rm enc}=2.85 ~\Gyr$. Fig. \ref{fig:NEuse} displays the energy distributions for all particles (upper panels) and for stellar particles (lower panels) in two models, Models PAR (left panels) and PIR (right panels). These models represent tangentially anisotropic and isotropic velocity dispersions for the intruder galaxy, respectively. At the initial state, the energy distribution of the particles is concentrated in a narrow range, as shown by the blue line. As the intruder galaxy falls inward, there is a redistribution of particle energies. At the moment of encounter, the energy range expands significantly, as shown by the cyan line. At $T_{\rm enc}=0.1\Gyr$, the dwarf galaxy undergoes expansion, with an increase in the number of escaping particles. The energy distribution of the bound particles (green curves) exhibits minimal changes. Thereafter, up to 2.85 Gyr, the energy distribution (orange curves) continues to evolve with some minor changes. This is because the intruder galaxy remains in a non-equilibrium state. The continuous expansion of the intruder galaxy coupled with the gradual formation of shells further cause particle energy evolution, although relatively small.

From the bottom panel of Fig. \ref{fig:NEuse}, we find that for the semi-analytical model with the same parameters of Model PAR (presented with gray solid line), the differential energy distribution is consistent with the N-body simulation (cyan solid line) at $T_{\rm enc}=0~\Gyr$. The deviation of differential energy between these two methods at low energy may come from the deformation of the intruder galaxy during the encounter in the N-body simulation. 

The differential energy distributions of these two methods are also compared with a ``velocity superposition approach'', defined as follows: take the initial phase space distribution of particles within the intruding galaxy at the beginning of the merger, place it at the presumed point of disruption, and simply add the velocity vector of the CoM of the intruding galaxy (drawn from the disruption point) to the velocity vectors of all particles, and compute the energy distribution from there. The gray dashed line in Fig. \ref{fig:NEuse} shows the differential energy from such superposition approach. The energy range of the particles is significantly broadened. However, compared to the N-body simulation, the energy distribution is dominated by a long tail of escape particles in the superposition approach. This long tail is a robust feature even if we adjust the exact value of the intruder CoM velocity by redefining the exact epoch of the disruption. In short, the superposition approach does not capture the physics as well as our semi-analytical method.

\subsection{Predicting number of observable shells at a given time}
\label{mergertime}
When the particles become unbound due to the disruption of the intruder galaxy in the free fall stage, their energies remain constant. The particles move within the gravitational field of a target galaxy. The energy range of the dwarf galaxy at the moment of disruption, as shown in Fig. \ref{fig:NEuse}, is crucial for limiting the number of observable shells at the given time. Radial periods $P_r$ is a function of energy $E_c$, as in the integration in Eq. \ref{radialE}, and is approximated as Keplerian as long as $r_{s1}<<r_{ap}$, where $r_{ap}$ is the orbital apocenter. 
Hence particles of different energies dictate different radial periods of motion by the following relation,
\begin{equation}
\label{eq17}
P_r(E)=\dfrac{2\pi GM_1}{\left(-2E\right)^{3/2}}, 
\end{equation}
where $P_r$ is defined as the time required for a star to travel from the apocenter to pericenter and return, and the energy $E$ of any particle within the $n_{\rm th}$ shell can be approximated by
\begin{equation}
\label{eq18}
E \approx \Phi_{1}(d_{n}),
\end{equation}
as shells are made of particles with nearly zero kinetic energy.
Here $d_{n}$ is the radius of the $n_{\rm th}$ shell. In the theoretical description of the shell system\citep{Quinn1984,hernquist1987}, the outermost shell is labeled as $n=1$. The outermost shell is always composed of stars that are reaching their apocenters for the second time. These particles have completed 3/2 oscillations, which include half an oscillation period from the point of release to the first apocenter passage. The shell labeled with index $n=2$ is composed of stars that are reaching their apocenters for the third time, and so on. While theory allows structures composed of particles to complete 1/2 radial oscillations, defining such structures as shells in N-body simulations has proven challenging. These structures appear as extended and diffuse envelopes in the projected density maps, devoid of distinct edges, making them difficult to observe.

There is a relationship between the time $T_{\rm enc}$ and the oscillation period $P_r(d_{n})$ of a particle within the $n_{\rm th}$ shell surrounding the center of a target galaxy:
\begin{equation}
\label{eq19}
T_{\rm enc} = \left(n+\tau_1-1\right)P_r(d_{n}), ~\tau_1 =3/2
\end{equation}
where $\tau_1$ is the count of oscillations for particles in the outermost shell.

From Eqs. \ref{eq17} and \ref{eq19}, we can derive the following equation,
\beq\label{eqnum} \nu=\frac{T_{\rm enc}(-2E_{min})^{3/2}}{2\pi GM_1} -\frac{1}{2}. \eeq
At a given time $T_{\rm enc}$, the shell formed by particles with energy $E_{\rm min}$ corresponds to the maximum shell label, $\nu$. The values of the minimum and maximum energies of particles, $E_{\text{min}}$ and $E_{\text{max}}$, can be determined in the analysis in  Section \ref{sec:Energydistribution}. In our semi-analytical model, once the intruding galaxy is disrupted, the energies of released particles remain constant. According to Eq. \ref{eqnum}, the number of observable shells is limited by the energy spread of particles at a given time $T_{\text{enc}}$. We calculated that at the given time $T_{\text{enc}}=2.85~\Gyr$, the values of $\nu$ in radial and near-radial merger models (using the same parameters as Models PAR, 1PANR and 3PANR) are $4.8$, $5.1$ and $6.2$, respectively. Compared to the N-body simulations, the second-generation shells are absent. This is due to the simplified assumption that the dwarf galaxy is disrupted during the first encounter in our semi-analytic model. In the future, we will further refine the semi-analytic model by considering the second and third generations of shells to obtain a more accurate energy range.

\section{Discussions and conclusions}
\label{discussion}
\subsection{Dynamical friction, energy evolution and mass models}
We employ semi-analytical calculations and simulations to investigate the process of shell formation, with a focus on the change in the distribution of particle energies within the intruder galaxy. During the infall process, the phenomenon of energy transfer among the particles in the intruder galaxy was initially reported and examined through a sequence of simulations conducted by \citet{Heisler1990}. Our study delves into the specifics of the energy redistribution process, and proposes and validates semi-analytical methods to quantify this energy evolution. We emphasize that the self-gravity of dwarf galaxies significantly expands the energy range of particles, and then affects the radial range of the shell, even when dynamical friction is not taken into account. We postpone a detailed study of dynamical friction, presently we find that dynamical friction is unimportant at the 10$\%$ level for our mass ratios M2:M1.

Our modeling of the particle energy evolution within the intruder galaxy is still worthy of further consideration, as we have chosen a model of the intruding galaxy that ensures complete or almost complete disruption during the first infall. If the intruder galaxy does not disrupt entirely in a single event, it would generate successive generations of shells during each subsequent pericenter passage, with each generation having its own energy distribution. Additionally, the gravitational potential changes of both the intruder galaxy ${\partial \Phi_2}/{\partial t}$ and the target galaxy ${\partial \Phi_1}/{\partial t}$ during the gradual disruption of the intruder may also play a role, to varying degrees, in the particle energy evolution. Modeling multiple generations of shells, as well as the effects of ${\partial \Phi_1}/{\partial t}$ and ${\partial \Phi_2}/{\partial t}$, would require significant time and effort and is out of the scope of this work. Currently, we emphasize the importance of the intruder galaxy's self-gravity for the energy spread of particles, and provide a semi-analytical calculation for the energy evolution during the first infall of the intruder galaxy.

At present, the potentials of the intruder galaxy and the target galaxy are all single-component Plummer/Hernquist model, in which the stellar particles of the intruder galaxy are treated as the most bound particles. The use of a multi-component galaxy model to stand for the intruding galaxy will have a different specific form of gravitational potential in our Eq. \ref{eq5}. However, the dark matter particles and stellar particles all follow the same role of gravity. They are both dissipationless particles. Thus a single-component galaxy model is sufficient enough to investigate the evolution of energy in the shell formation. We will consider multi-components (dark matter and stars) and a realistic NFW model in our future work.

\subsection{Conclusions}
\label{summary}
Our work finds analytical formulae for the energies and angular momenta of particles of a spherical anisotropic dwarf galaxy, as described by Eqs. \ref{eq5} and \ref{eq6}, gained between its infall and its disruption in a massive elliptical. During the moment of disruption, particles that are released from the ``front seats'' of the dwarf galaxy experience a loss of energy, while those released from the ``back seats'' gain energy (cf. Fig. \ref{fig:CCWCW}). The evolution of angular momenta mimics that of energies but lags by a phase of $\frac{\pi}{2}$. These surprisingly neat results hold true both for isotropic and for anisotropic dwarfs on nearly radial infall, as the internal kinetic energy of the dwarf is a negligible budget in the energy evolution.

The energy range for stellar particles in the progenitor in our model is significantly wider compared to Quinn's (1984) model. To match this energy range for stellar particles within an intruding galaxy, Quinn's initial model would require a higher mass for this galaxy. Moreover, this expanded energy range results in a higher shell count in our model. When estimating the mass of the progenitor by counting the shells in an observed shell galaxy, Quinn's model tends to overestimate the mass.

Utilizing our semi-analytical model, we calculate the energy range of particles within the intruder galaxy when the intruder galaxy arrives at the pericenters along different orbits. We predict the number of shells above the minimum energy $E_{\rm min}$ at a given time after the first encounter between two galaxies, which is close to the result of N-body simulation, although the second-generation shells are not taken into account.

\section*{ACKNOWLEDGEMENTS}
The authors acknowledge the support of the National Key Research and Development Program of China (2023YFA1608100). X.W. wishes to thank the financial support from the Natural Science Foundation of China (Number NSFC-12073026) and ``the Fundamental Research Funds for the Central Universities'' (KY2030000173, WK3440000004), which made this research possible. H.Z. acknowledges the UK Science and Technology Facilities Council grant ST/V000861/1 and the USTC Fellowship for International Cooperation for the funding provided. L.F. gratefully acknowledges the National Natural Science Foundation of China (NSFC, grant No. 12173037, 12233008), the CAS Project for Young Scientists in Basic Research (No. YSBR-092) and the Fundamental Research Funds for the Central Universities (WK3440000006). All the authors thank Cyrus Chun Ying Tang Foundations and the 111 Project for ``Observational and Theoretical Research on Dark Matter and Dark Energy'' (B23042). 

\bibliographystyle{aasjournal}
\bibliography{ref} 

\end{document}